\documentclass[10pt]{article}
\usepackage[utf8]{inputenc}
\usepackage{booktabs}
\usepackage{tabularx}
\usepackage{graphicx}
\usepackage{hyperref}
\graphicspath{ {./} }
\usepackage[T1]{fontenc}
\usepackage{float}
\usepackage[super]{natbib}

\usepackage{amssymb,amsmath,amsfonts,eurosym,geometry,ulem,graphicx,caption,color,setspace,sectsty,comment,footmisc,caption,natbib,pdflscape,subfigure,array,hyperref}

\newcolumntype{L}[1]{>{\raggedright\let\newline\\arraybackslash\hspace{0pt}}m{#1}}
\newcolumntype{C}[1]{>{\centering\let\newline\\arraybackslash\hspace{0pt}}m{#1}}
\newcolumntype{R}[1]{>{\raggedleft\let\newline\\arraybackslash\hspace{0pt}}m{#1}}

\begin{document}

\begin{titlepage}
\title{Measuring Racial Disparities in Rent Growth Under Algorithmic Landlord Concentration in U.S. Metros}
\author{
\textbf{Advay Ranade}\thanks{Independent Researcher, Sunnyvale, CA. Research Assistant, Stanford University Deliberative Democracy Lab, CA. Corresponding Author, Project Lead. \textbf{Email: advaymranade@gmail.com} }}
\date{June 2026}
\maketitle
\begin{abstract}
\noindent The 2024 Department of Justice antitrust complaint against RealPage, Inc. named five major residential REITs for coordinating algorithmic rent pricing across hundreds of thousands of apartment units in major US metropolitan areas. This paper studies whether census-tract-level corporate landlord concentration (CLC), measured from SEC EDGAR 10-K property filings geocoded to census tracts, the first such application in the literature, is associated with rent growth 2019-2023, and whether that association is larger in majority-minority neighborhoods. Rent outcomes are measured using the Zillow Observed Rent Index (ZORI). To account for the possibility that corporate landlords preferentially locate in neighborhoods already seeing rent appreciation, all regressions control for a fully novel Algorithmic Housing Burden Index (AHBI), a composite of pre-existing rent burden and market tightness from ACS data. Across 665 census tracts in ten US metropolitan areas, doubling REIT concentration is associated with 2.8 percentage points higher rent growth (p = 0.086, p = 0.030, HC1 robust). This association is significantly stronger in majority-minority tracts. Within the same metro, high-CLC majority-minority tracts are associated with 5.9 percentage points higher rent growth than comparable white tracts (p = 0.039). An XGBoost model predicts 44 percent of out-of-sample rent growth variance, with SHAP analysis independently confirming that CLC's contribution is positive in minority tracts and negative in white tracts. Taken all together, these findings provide the first tract-level evidence consistent with corporate landlord concentration being associated with disproportionately higher rent growth in communities of color.

\end{abstract}
\end{titlepage}

\section{Introduction} \label{sec:introduction}
The rapid adoption of algorithmic rent-pricing software by corporate landlords has created a new, largely unexamined mechanism through which housing inequality in the United States is actively widening. These systems, most prominently RealPage's YieldStar platform, coordinate pricing decisions across competing landlords at scale, processing real-time occupancy and marketing data to generate rent recommendations that individual property managers are strongly incentivized to follow. In 2024, the United States Department of Justice filed a civil antitrust complaint naming five major residential real estate investment trusts (REITs)\cite{doj2024}:
\begin{itemize}
  \item Avalon Bay Communities
  \item Equity Residential
  \item Essex Property Trust
  \item Mid-America Apartment Communities
  \item UDR
\end{itemize}
These REITs were named for their participation in this coordinated pricing scheme. The complaint alleges that these firms, which collectively own hundreds of thousands of apartment units across major U.S. metropolitan areas, used RealPage software to suppress competitive rent settings in ways that harmed their tenants.

Existing research establishes the aggregate rent effects of this coordination. Calder Wang and Kim (2024)\cite{calderwang2024} show that, using a differences-in-differences design, markets where landlords adopt algorithmic pricing exhibit rents approximately 4 percent higher than unpenetrated markets, with the markup reflecting joint profit maximization rather than independent competitions in a standard capitalist economy. The White House Council of Economic Advisers (2024)\cite{cea2024} estimates that this rent inflation cost American renters 3.8 billion dollars in 2023 alone, noting that the harm was amplified where housing supply is constrained. Additional work by the UC Berkeley School of Information (2024)\cite{majidzadeh2024} documents substantial rent increases following RealPage adoption at the property level. However, none of these studies examine whether the distributional burden of algorithmic rent pricing falls equally across racial groups or whether the concentration of corporate landlords at the neighborhood level generates racially heterogeneous rent outcomes. This is the gap this research study addresses.

We study whether census tract-level corporate landlord concentration, defined as the share of renter-occupied units owned by the five DOJ-named REITs, can predict artificial rent growth between 2019 and 2023. We also study whether that effect is disproportionately concentrated in majority-minority census tracts. CLC is constructed from SEC EDGAR 10-K Schedule III filings from fiscal years 2016 through 2022, which require REITs to disclose their full property portfolios with addresses and unit counts. We geocode the property names and cities to census tracts, yielding a tract-level measure for algorithmic landlord exposure built entirely on public regulatory data. Rent outcomes are measured using the Zillow Observed Rent Index (ZORI), a transaction-based monthly rent dataset at the zip code level that captures the 2021-2022 rent surge without the smoothing effects of ACS survey data. We additionally construct the Algorithmic Housing Burden Index (AHBI), a composite of pre-existing rent burden and market tightness, which serves both to characterize pre-existing vulnerability in high CLC tracts and as a control variable separating CLC's marginal effect from baseline market conditions.

We find evidence consistent with both hypotheses created in the present paper. For H1, doubling REIT concentration in a census tract is associated with approximately 2.8 percentage points higher rent growth over 2019-2023, controlling for baseline housing stress and demographic composition (p = 0.086, metro-clustered standard errors; p = 0.030, HC1 robust standard errors). For H2, we find that the CLC effect on rent growth is significantly larger in majority-minority census tracts than in majority-white tracts within the same metropolitan area. Within metros, tracts with high CLC and majority-minority composition exhibit 5.9 percentage points higher rent growth than comparable majority-white tracts (p = 0.039). This disparity is confirmed using a continuous minority share interaction, which is significant at p = 0.002. For H3, an XGBoost model trained on 2019 tract-level features predicts 44 percent of the variance in rent growth on tracts excluded from training data. CLC ranks fifth out of eight features by SHAP analysis, independently confirming the H2 racial disparity: CLC’s predicted contribution to rent growth is positive in minority tracts and negative in white tracts with the OLS interaction results. 

Overall, this paper makes four major contributions. First, we introduce a publicly replicable pipeline for measuring algorithmic landlord concentration at the census tract level from SEC EDGAR filings, a data source that has never been previously used for a neighborhood-level housing analysis. Second, we construct the AHBI, a novel, composite index of housing vulnerability from ACS public data, providing a generalizable baseline characterization tool for future research on this topic. Third, we provide among the first tract-level estimates of racial disparities in rent growth attributable to corporate landlord concentration, using OLS interaction models as well as non-parametric machine learning validation. Fourth, we document a methodological finding with implications for future housing research. This is that ACS 5-year estimates are insufficient to detect actual rent growth due to multi-year smoothing, and transaction-based data sources such as ZORI are necessary for this type of analysis.

\section{Literature Review} \label{sec:literature}
The pertinent literature on this study falls into three distinct areas of research that have, prior to this study, developed separately. The first area of study documents the rent-increasing effects of algorithmic rent pricing in all types of rental markets. The second establishes the persistence and structural roots of racial and income-based disparities in housing cost burden and homeownership. The third examines how algorithmic tools in similar portions of the housing market have produced racially different outcomes at the census-tract level. Each strand makes meaningful contributions to understanding housing inequality in the United States. However, no study has drawn them all together to analyze whether algorithmic rent pricing disproportionately burdens high-minority census tracts or widens the racial homeownership gap at the census-tract level. This paper directly addresses that gap.

\subsection{Evidence of Past Algorithmic Rent Inflation}
Since 2022, research on algorithmic rent pricing has expanded rapidly, driven in large part by the substantial increase in revenue management products on the market, notably AIRM, formerly YieldStar. The existing literature on this topic establishes that algorithmic pricing raises rents above where prices would be in a competitive market, but, studies differ in the estimates of magnitude and in the mechanisms that are identified. The most rigorous study in this topic is Calder-Wang and Kim (2024)\cite{calderwang2024}, who hand-collect algorithmic pricing adoption decisions across a panel of market-rate family buildings between 2005 and 2019. This data is then merged with building-level rent and occupancy records from Moody’s REIS across fifty metropolitan areas. Using a differences-in-differences design with fixed effects\cite{callaway2021} as well as robustness checks with building an algorithmic pricing adoption estimator, they document that algorithmic adopters set prices more responsively to market conditions as well as at higher rates than non-adopters. More directly connected to this analysis, other testing from past literature finds that observed markups among same-software users are more consistent with joint profit maximization than with competition, contradictory to a competition-based free market. Due to the focus of joint profit maximization, a fully algorithmically-penetrated market exhibits average rents of approximately four percent higher than an unpenetrated market during specific time periods\cite{calderwang2024}.

The White House Council of Economic Advisers (2024)\cite{cea2024} extends past work using more recent data and estimates that algorithmic pricing cost renters an average of 70 dollars per month in 2023, totaling 3.8 billion dollars nationally. The same cost-effectiveness analysis report furthers that harm will be larger in areas where housing supply is more constrained and renters have less financial cushion, This specific observation directly provides the empirical basis for the pre-existing rent burden component within the Algorithmic Housing Burden Index (AHBI), which is developed in this study. Another investigation done by the UC Berkeley School of Information (2024)\cite{majidzadeh2024} uses propensity score matching and machine learning methods to confirm a statistically significant rent premium associated with RealPage usage, while differences-in-differences analyses of the 2017 RealPage acquisition of Lease Rent Options, found no market-wide rent increase directly connected to the merger\cite{calderwang2024}. Together these three studies establish a well-supported consensus that algorithmic pricing software raises rents measurable above competitive levels. However, whether the rent-increasing effect falls disproportionately on minority communities and compounds racial gaps in homeownership remains unknown until this paper.

\subsection{Structural Foundations of Racial Housing Disparities}
Many studies document persistent racial disparities in housing cost burden, homeownership rates, and housing wealth accumulation in the United States. Using fifty years of longitudinal survey data, Colburn et al. (2024)\cite{colburn2024} find that the probability of Black households experiencing housing cost burden is close to 50 percent higher than for comparable White households after controlling for income, neighborhood characteristics, and metropolitan context. This finding directly motivates the inclusion of pre-existing rent burden as a component of the AHBI because tracts where renters are already financially stretched represent populations with the least capacity to absorb algorithmic price increases. Additionally, the racial gap is also present in homeownership studies. The U.S. Treasury Department (2022)\cite{treasury2022} documents a gap of approximately 30 percentage points between White and Black homeownership rates that has changed close to nothing over 30 years despite the Fair Housing Act of 1968. Brookings Institution (2024)\cite{greene2024} furthers that at the census-tract level, homes in Black-majority neighborhoods are valued at median prices close to four times lower than comparable homes in Black-minority neighborhoods even though both communities reside in the same metropolitan area. Additionally, 17 percent of the racial homeownership gap cannot be explained by observable demographic factors\cite{urbaninstitute2024}, and, the racial wealth gap between median Black and White households essentially unchanged from 1992 to 2022, with the housing channel accounting for a substantial share of that stagnation\cite{ncrc2024}. The existing research in this area is comprehensive in documenting the existence and magnitude of racial housing disparities, but does not identify algorithmic pricing as a mechanism through which these disparities are widening. The present study addresses this limitation directly.

\subsection{Racial Bias in Algorithmic Real Estate Pricing}
A third and more recent area of research examines how algorithmic tools in similar segments of the housing market have produced racially unequal outcomes at the neighborhood level, providing both theoretical motivation and precedent for the present study. Raymond (2024)\cite{raymond2024} finds that algorithmic valuation models used by corporations such as Zillow or Redfin to estimate the prices of homes, can compound already disproportionate effects on housing values in Black-majority neighborhoods, with further implications for the racial homeownership gap. The Urban Institute (2020)\cite{urbaninstitute2023} quantifies racial inequality in the accuracy of the automated valuation model at the census tract level, finding that these models systematically undervalue homes in Black-majority neighborhoods relative to comparable homes in white-majority neighborhoods. This pattern is directly attributable to systemic discrimination in the training data of automated valuation models. This finding establishes a critical precedent that algorithmic real estate tools can and do produce racially heterogeneous outcomes at the census-tract level, measurable using publicly available data. 

\subsection{Synthesis and the Research Gap}
Taken together, these three areas of research establish that algorithmic rent pricing raises rents measurably above competitive levels in algorithmic landlord concentrated markets, that racial and income disparities in housing cost burden and homeownership are persistent, large, and structurally rooted, and that algorithmic tools in adjacent housing markets produce racially disparate outcomes at the census-tract level. However, no existing study has connected these three bodies of evidence, measuring whether census-tract-level exposure to corporate algorithmic landlords can correlate with the acceleration of the racial homeownership gap and income-based rent burden, and whether this effect is larger in minority tracts than in white tracts. This study addresses that gap by constructing a novel Algorithmic Housing Burden Index (AHBI) at the census-tract level across the ten largest metropolitan areas in the United States and using it as the primary treatment variable in a difference-in-differences analysis of homeownership and rent burden outcomes between 2019, 2023, and in the future. The contributions of this study specifically include the AHBI, a novel index that is the first replicable, publicly sourced measure of algorithmic pricing exposure at the census tract level. The AHBI can provide the first causal estimates of the racial channel through which algorithmic pricing may compound existing inequalities and the causal identification can also be paired with a predictive XGBoost model, rather than propensity score matching like other research. This would allow for anyone to be able to identify the tract characteristic most predictive of algorithmic pricing harm, helping provide a foundation for future policy intervention and implications.

\section{Methodology} \label{sec:method}
\subsection{Conceptual Framework}
This paper studies whether corporate landlord concentration causes rent growth between 2019 and 2023, and whether that effect falls disproportionately on minority census tracts. The central identification challenge is that corporate landlords cluster in markets with specific pre-existing characteristics such as urban density, high existing rents, and constrained supply. These other factors may allow us to independently predict future rent increases. Any empirical strategy must therefore separate the effect of corporate landlord presence from the baseline conditions that attracted it.

We address this through a three-construct design. The first is Corporate Landlord Concentration (CLC), which measures the degree to which REIT-affiliated landlords control the rental stock within a census tract in 2019. The second is the Zillow Observed Rent Index (ZORI), a transaction-based measure of observed rents at the ZIP code level, which we use to construct tract-level rent growth between 2019 and 2023 as the primary outcome variable. The third is the Algorithmic Housing Burden Index (AHBI), a composite of pre-existing rent burden and market tightness that is used as a characterization tool showing that high-CLC tracts were already more financially stressed prior to the study period as well as a control variable in the primary regressions.

\subsection{Corporate Landlord Concentration (CLC)}
\subsubsection{Data Source}
CLC is constructed from the SEC EDGAR 10-K annual filings of five major publicly traded residential real estate investment trusts (REITs): AvalonBay Communities (AVB), Equity Residential (EQR), Essex Property Trust (ESS), Mid-America Apartment Communities (MAA), and UDR, Inc. (UDR). These five REITs were named in the Department of Justice's 2024 civil antitrust complaint\cite{doj2024} regarding algorithmic rent coordination through RealPage's revenue management software, and collectively represent a significant share of institutionally owned multifamily rental housing in major US metropolitan markets.

Residential REITs are required to disclose their full property portfolios in Schedule III of their 10-K filings, including property names, addresses, and unit counts. We collect 10-K filings for fiscal years 2016 through 2022 from the SEC EDGAR full-text search system, yielding a panel of property-level observations spanning the study period. Property addresses are geocoded to latitude and longitude coordinates using the Google Maps Geocoding API and subsequently crosswalked to 2020 census tract GEOIDs using the Census Bureau's geocoder.

\subsubsection{Construction}
CLC for census tract i in year t is defined as:

$$\text{CLC}_{it} = \frac{\sum_{j \in i} \text{REIT units}_{jt}}{\text{Renter-occupied units}_{it}}$$
where the numerator is the sum of all REIT-owned units across properties located within tract i in year t, and the denominator is the total count of renter-occupied housing units in the tract from ACS Table B25003. CLC can exceed 1.0 in tracts where REIT unit counts that are drawn from property schedules that may include units under construction, exceed the ACS renter-occupied stock.

For properties with missing unit counts in a given filing year, we impute from the most recent prior year's filing for the same property. Properties for which no unit count can be recovered across any filing year are excluded from the numerator. We use 2019 CLC as the pre-period treatment measure throughout the analysis. Twelve census tracts are excluded from the CLC where 2019 ACS renter-occupied units fall below 150, concentrated in rapidly developing suburban areas of Dallas-Fort Worth and Charlotte where REIT construction activity outpaced the ACS's ability to measure renter-occupied stock. These tracts are retained in all other analyses.

\subsubsection{Winsorization and Transformation}
The raw CLC distribution is highly right-skewed, with a small number of tracts recording values exceeding 10. These cases are when REIT unit counts substantially exceed ACS-measured renter-occupied units, typically due to new, rapid construction activity in developing suburban tracts where the ACS stock count lags behind actual occupancy. We winsorize CLC at the 99th percentile of the analysis sample before estimation. We additionally apply a log transformation, $\log(1 + \text{CLC})$, as the primary specification, so that the regression coefficient reflects the effect of a proportional increase in corporate landlord presence rather than a unit increase. Results using raw winsorized CLC are reported as a robustness check.

\subsection{Baseline Vulnerability Characterization: Algorithmic Housing Burden Index}
To describe the housing stress conditions present in high-CLC tracts before the study period, we construct the Algorithmic Housing Burden Index (AHBI), a composite of two tract-level measures drawn from ACS 5-year estimates. The AHBI serves two roles in the analysis: as a descriptive tool characterizing the pre-existing vulnerability of high-CLC tracts, and as a control variable ($\text{AHBI}_{\text{pre}}$) in the primary regression, which separates CLC's effect on rent growth from the baseline market conditions that attracted corporate landlords to those tracts in the first place.

\subsubsection{Pre-existing Rent Burden (PRB)}
PRB is defined as the share of renter households in a census tract paying 30 percent or more of gross income on rent, drawn from ACS Table B25070:
$$\text{PRB}_{it} = \frac{\sum_{k \geq 30\%} \text{renters in burden bracket}_{kit}}{\text{Rent burden universe}_{it}}$$
This measure captures the financial fragility of the renter population. Households already at or above the conventional affordability threshold have minimal buffer to absorb further rent increases driven by algorithmic coordination.

\subsubsection{Housing Market Tightness (HMT)}
HMT is defined as the complement of the vacancy rate in a census tract, computed from ACS Table B25002:
$$\text{HMT}_{it} = 1 - \frac{\text{Vacant units}_{it}}{\text{Total housing units}_{it}}$$
Low vacancy means renters facing rent increases have few viable alternatives within the local market. Tight markets eliminate the competitive discipline that individual mobility would otherwise impose on landlord pricing decisions, amplifying the effect of algorithmic coordination.

\subsubsection{AHBI Construction}
Each component is standardized as a z-score relative to the cross-tract distribution in the 2022 reference year and averaged equally:
$$\text{AHBI}_{it} = \frac{1}{2} \cdot \frac{\text{PRB}_{it} - \bar{\text{PRB}}_{2022}}{\sigma_{\text{PRB},2022}} + \frac{1}{2} \cdot \frac{\text{HMT}_{it} - \bar{\text{HMT}}_{2022}}{\sigma_{\text{HMT},2022}}$$
Equal weighting reflects the theoretical symmetry between the two dimensions. Principal component analysis was evaluated as a data-driven weighting alternative. CLC loaded almost entirely on the third principal component (15 percent of variance), confirming its statistical independence from PRB and HMT and motivating its treatment as an external variable rather than an index component.

\subsection{Primary Outcome: Zillow Observed Rent Index (ZORI)}

\subsubsection{Why ZORI Rather than ACS Rent Measures}
ACS-based rent burden estimates are 5-year survey averages that smooth over any single year's conditions. The ACS 2023 5-year vintage covers survey years 2019-2023, averaging together the COVID-era rent softening of 2020-2021 and the acute national rent surge of 2021-2022. This smoothing makes ACS incapable of detecting the rent spike that is the central outcome of interest in this study. This was found when empirical tests using $\Delta$ AHBI and $\Delta$ homeownership gap as outcomes produced null and inconsistent results across multiple ACS vintage combinations, consistent with this measurement limitation.

The Zillow Observed Rent Index (ZORI) is a transaction-based rent index constructed from observed asking rents for rental units listed on Zillow's platform, smoothed and seasonally adjusted, published monthly at the ZIP code level. Because ZORI is built from actual rental listings rather than survey responses, it captures the rent surge as it occurred in real time with no multi-year averaging. ZORI is the standard data source in the empirical rent pricing literature, used in Calder-Wang and Kim (2024) and the White House CEA (2024) analyses among others.

\subsubsection{Construction and Zip-to-Tract Conversion}
We download the ZORI smoothed, seasonally adjusted series for all home types at the ZIP code level from Zillow's public research data repository, covering January 2014 through December 2023. Pre-period ZORI for each ZIP is computed as the average of the twelve monthly observations in calendar year 2019. Post-period ZORI is the average of the twelve monthly observations in calendar year 2023. Rent growth for ZIP code $z$ is:

$$\text{Rent Growth}_{z} = \frac{\bar{\text{ZORI}}_{z,2023} - \bar{\text{ZORI}}_{z,2019}}{\bar{\text{ZORI}}_{z,2019}}$$

Because our primary analysis operates at the census tract level, we crosswalk ZIP-level ZORI to census tracts using the HUD USPS ZIP-to-Tract crosswalk (Q4 2019), which provides residential address allocation factors $w_{zt}$ representing the fraction of ZIP code $z$'s residential addresses that fall within tract $t$. Tract-level rent growth is computed as a weighted average across all ZIP codes that overlap the tract:

$$\text{Rent Growth}_{t} = \frac{\sum_{z} w_{zt} \cdot \text{Rent Growth}_{z}}{\sum_{z} w_{zt}}$$

Of the 689 tracts in the analysis sample, 583 (84.6 percent) are matched to at least one ZIP with complete 2019 and 2023 ZORI observations. Unmatched tracts are primarily in lower-density suburban areas with insufficient rental listing activity on Zillow's platform, and are excluded from the ZORI regression sample.

\subsubsection{Spatial Autocorrelation Complication}
Because multiple census tracts can fall within a single ZIP code, tracts sharing a ZIP are assigned identical or near-identical rent growth values by construction. This mechanical overlap induces positive spatial autocorrelation in the regression residuals (Durbin-Watson statistic = 0.46 in the H1 specification without metro fixed effects). We address this in two ways: first, by reporting metro-clustered standard errors that account for within-metro error correlation; second, by noting this as a study limitation and encouraging replication at the ZIP code level, which eliminates the mechanical correlation. The H2 specification, which includes metro fixed effects, exhibits substantially less spatial autocorrelation (Durbin-Watson = 0.81) and is our preferred specification for inference about the racial disparity channel.

\subsection{Regression Specifications}
\subsubsection{H1: Average Effect of CLC on Rent Growth}
The hypothesis denoted as H1 estimates the average relationship between 2019 CLC and tract-level rent growth between 2019 and 2023:
$$\text{Rent Growth}_{i} = \alpha + \beta_1 \log(1 + \text{CLC}_{i,2019}) + \beta_2 \text{AHBI}_{i,2019} + \mathbf{X}_i'\boldsymbol{\gamma} + \varepsilon_i$$
where $\mathbf{X}_i$ is a vector of 2019 tract-level controls (median household income, renter share, percent Black, percent Hispanic, total housing units), and standard errors are clustered at the metropolitan area level. Metro fixed effects are excluded from this specification to preserve between-metro variation, which is central to identification: high-CLC Sun Belt metros (Dallas-Fort Worth, Charlotte, Atlanta) exhibited substantially higher rent growth than low-CLC coastal metros (San Francisco, San Jose), and restricting comparison to within-metro variation eliminates this signal. The coefficient of interest is $\beta_1$: a positive estimate indicates that higher corporate landlord concentration predicts greater rent growth.
The inclusion of $\text{AHBI}_{i,2019}$ as a control is essential to this specification. AHBI is itself significantly positively associated with rent growth ($\beta_2 = +0.176$, $p < 0.001$), reflecting that pre-existing tight, high-burden markets continued to see above-average rent growth. Controlling for AHBI separates CLC's marginal effect from these baseline conditions, addressing the non-random location of corporate landlords.

\subsubsection{H2: Varying Effects of Tract Racial Comparison}
The H2 specification adds metro fixed effects and tests whether the CLC effect on rent growth is larger in majority-minority tracts:

\begin{equation}
\begin{aligned}
\text{Rent Growth}_{i} = \; & \alpha + \beta_1 \log(1 + \text{CLC}_{i,2019}) + \beta_2 \text{Minority}_{i} \\
& + \beta_3 \left[ \log(1 + \text{CLC}_{i,2019}) \times \text{Minority}_{i} \right] \\
& + \beta_4 \text{AHBI}_{i,2019} + \mathbf{X}_i'\boldsymbol{\gamma} + \delta_m + \varepsilon_i
\end{aligned}
\end{equation}

where $\text{Minority}_{i}$ is an indicator equal to 1 if the combined Black and Hispanic share of tract $i$'s population exceeds 50 percent, and $\delta_m$ are metropolitan area fixed effects. Standard errors are clustered at the metro level. The coefficient of interest is $\beta_3$: a positive estimate indicates that CLC has a disproportionately large effect on rent growth in majority-minority tracts relative to majority-white tracts in the same metropolitan area.
We additionally test H2 using a continuous minority share variable interacted with log CLC, replacing the binary $\text{Minority}_{i}$ indicator to confirm the result does not depend on the choice of threshold.

\subsection{H3: XG Boost Prediction Model}
\subsubsection{Purpose and Relationship}
The OLS specifications in Sections 3.5.1 and 3.5.2 estimate average and heterogeneous effects of CLC on rent growth under linearity and selection-on-observables assumptions. As a complementary exercise, we trained an XGBoost model that serves two purposes. First, it tests whether tract-level algorithmic landlord exposure has out-of-sample predictive power for rent growth beyond what demographic and market controls can suggest. Second, SHAP (SHapley Additive exPlanations) values computed from the trained model provide a non-parametric check on the H2 racial disparity finding. SHAP checks if CLC’s contribution to predicted rent growth is systematically positive in minority tracts and negative in white tracts. 

\subsubsection{Feature Set and Target Variable}
The model is trained on eight 2019 tract-level features: $\log(1 + \text{CLC}_{i,2019})$, $\text{AHBI}_{i,2019}$, percent Black, percent Hispanic, percent white non-Hispanic, renter share, median household income, and total housing units, The target variable is tract-level rent growth in years 2019-2023 as defined in Section 3.4.2 The feature set is intentionally identical to the controls used in the OLS specifications, ensuring that any predictions attributable to $\log\text{CLC}$ reflect the same variation being studied in other parts of the study. Critically, demographic features are included so that the model can express the racial interaction. This allows SHAP decompositions to show whether CLC's marginal contribution differs by tract racial composition.

\subsubsection{Model Training and Hyperparameter Tuning}
The 583-tract sample with complete ZORI coverage is randomly partitioned into a training set (80 percent, n = 466) and a held-out set (20 percent, n = 117). All hyperparameter selection is conducted exclusively on training data using 5-fold cross-validation to minimize data leakage. The test set is used exactly once, solely for final evaluation. The hyperparameter search covers the following grid: learning rate $\in \{0.01, 0.05, 0.1\}$, maximum tree depth $\in \{3, 4, 5\}$, subsample fraction $\in \{0.6, 0.8, 1.0\}$, column subsample fraction $\in \{0.6, 0.8, 1.0\}$, minimum child weight $\in \{1, 3\}$, and L2 regularization $\in \{0.1, 1, 5\}$. Early stopping is applied within each cross-validation fold, halting training when validation RMSE fails to improve for 30 consecutive rounds. The selected hyperparameters are reported in Table 1.

\begin{table}[H]
\centering
\caption{Hyperparameter Final Selected Values}
\label{tab:tab2}
\renewcommand{\arraystretch}{1.5}
\begin{tabularx}{\textwidth}{XX} 
\toprule
\textbf{Hyperparameter} & \textbf{Selected Value} \\
\midrule
Learning Rate & 0.05 \\
Max tree depth & 4 \\
Subsample fraction & 0.6 \\
Column subsample fraction & 0.8 \\
Minimum child weight & 1 \\
L2 regularization ($\lambda$) & 1 \\
Number of estimators & 94 \\  
CV RMSE & 0.0913 \\  
\bottomrule
\end{tabularx}
\end{table}

The final model achieves a training $R^2 = 0.745$ and a test $R^2 = 0.439$, indicating moderate out-of-sample predictive power for a neighborhood-level problem. The train-test gap of 0.306 is expected given the regularization applied and the small sample ($n = 583$), but the test $R^2$ of 0.439 represents substantial explanatory power, nearly half the variance in 2019-2023 rent growth across held-out tracts is predicted from 2019 tract characteristics alone.

\section{Summary Statistics} \label{sec:summary}
\subsection{Sample Overview}
The analysis sample comprises 665 census tracts across ten metropolitan areas: Atlanta, Charlotte, Dallas–Fort Worth, Los Angeles, New York, San Diego, San Francisco, San Jose, Seattle, and Washington DC. Of these, 583 tracts (87.7 percent) are matched to at least one ZIP code with complete 2019 and 2023 Zillow ZORI observations and form the regression estimation sample; the remaining 82 tracts are retained for descriptive analysis but excluded from the rent growth regressions due to insufficient rental listing activity in Zillow's platform. The five study REITs collectively own 348,121 units across 980 properties within the sample, representing 18.6 percent of the total housing stock across all 665 tracts.
Of the 665 tracts, 59 (8.9 percent) are classified as majority-minority (combined Black and Hispanic share exceeding 50 percent) and 606 (91.1 percent) as majority-white. Eighty-five tracts (12.8 percent) record zero REIT unit presence in 2019 and therefore have $\log(1 + \text{CLC})=0$ the remaining 580 tracts have at least some corporate landlord exposure. Table 2 reports summary statistics for the full sample and separately for majority-minority and majority-white tracts.

\begin{table}[H]
\centering
\caption{Overall Analysis Variables}
\label{tab:tab2}
\renewcommand{\arraystretch}{1.5}
\begin{tabularx}{\textwidth}{XXXXXX} 
\toprule
\textbf{Variable} & \textbf{N} & \textbf{Mean} & \textbf{SD} & \textbf{Min} & \textbf{Max} \\
\midrule
Rent Growth 2019-2023 & 583 & 0.191 & 0.120 & -0.143 & 0.461 \\
Log(1+CLC) & 665 & 0.314 & 0.382 & 0.000 & 2.241 \\
AHBI (2019) & 665 & 0.568 & 0.135 & 0.121 & 0.937 \\
Median Household Income & 665 & 98,119 & 36,097 & 17,237 & 250,001 \\
Renter Share & 665 & 0.609 & 0.231 & 0.012 & 1.000 \\
Minority Share & 665 & 0.233 & 0.168 & 0.012 & 0.983 \\
\bottomrule
\end{tabularx}
\end{table}

\subsection{Outcome Variable: Rent Growth}
Tract-level rent growth between 2019 and 2023 averages 19.1 percentage points (SD = 12.0 pp) across the ZORI-matched sample, with considerable dispersion (interquartile range: 10.9 pp to 28.1 pp). The distribution is broadly symmetric around the mean but with a slight right skew driven by high-growth Sun Belt markets. At the metro level, rent growth is highest in San Diego (mean 38.2 pp), Charlotte (32.1 pp), Atlanta (28.3 pp), and Dallas-Fort Worth (28.6 pp), and substantially lower in San Francisco (4.3 pp) and San Jose (6.9 pp). This between-metro variation, which spans more than 30 percentage points from top to bottom, reflects the combination of supply constraints, post-COVID migration patterns, and differential corporate landlord penetration across markets, and constitutes the primary source of identifying variation for the H1 specification.

Majority-minority tracts experienced meaningfully higher average rent growth (24.3 pp) than majority-white tracts (18.7 pp), a raw gap of 5.6 percentage points. This unconditional gap motivates the H2 test, which asks whether this disparity persists after controlling for pre-existing market conditions and metro fixed effects and whether it is specifically attributable to CLC rather than other tract characteristics.

Figure 1 displays mean log(1 + CLC) and mean ZORI rent growth side by side for each metropolitan area, sorted by corporate landlord concentration from lowest to highest. The figure reveals the between-metro variation driving the H1 identification: Sun Belt markets (Dallas-Fort Worth, Charlotte, Atlanta) record both the highest mean CLC and above-average rent growth, while San Francisco and San Jose record the lowest CLC and the lowest rent growth in the sample. One notable exception to the pattern is San Diego, which records the lowest mean CLC among the ten metros yet the highest average rent growth (38.2 percent), reflecting supply constraints unrelated to institutional landlord penetration. This exception underscores why the regression analysis, which controls for metro-level factors, is necessary to isolate the CLC effect.

\begin{figure} [H]
    \centering 
    \includegraphics[width=12cm]{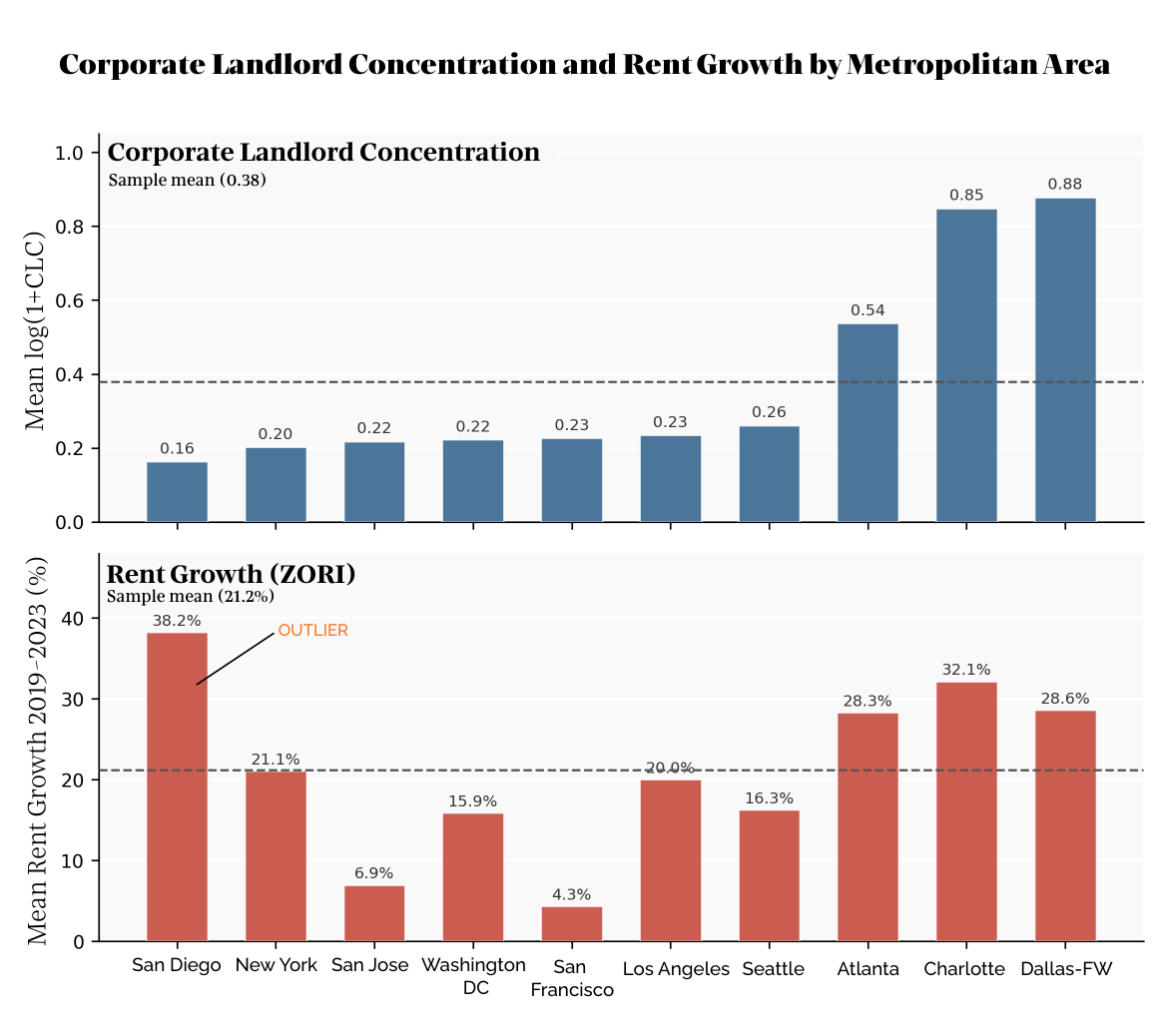}
    \caption{Mean corporate landlord concentration (log(1 + CLC), panel a) and ZORI rent growth 2019–2023 (panel b) by metropolitan area, sorted ascending by CLC. Sun Belt markets (Dallas–Fort Worth, Charlotte, Atlanta) lead on both measures; San Francisco and San Jose rank lowest. Source: SEC EDGAR 10-K filings and Zillow ZORI.}
    \label{fig:fig1_metro_clc}
\end{figure}

\section{Results} \label{sec:result}

\subsection{H1: Corporate Landlord Concentration and Rent Growth}
Table 3 reports OLS estimates of the H1 specification, which regresses tract-level ZORI rent growth on log(1 + CLC) and controls without metro fixed effects, using two standard error approaches. The primary coefficient on log CLC is +0.0281 (SE = 0.0164, p = 0.086 metro-clustered; p = 0.030 HC1 robust), implying that a doubling of REIT concentration within a census tract is associated with approximately 2.8 percentage points higher rent growth over the 2019–2023 period. Across 583 tracts, the model explains 28.7 percent of the variance in rent growth ($R^2$ = 0.287), with the large majority of variation attributed to between-metro differences in supply constraints and post-COVID migration patterns captured in the control variables.

The H1 result is directionally consistent across standard error assumptions but crosses conventional significance thresholds only under heteroscedasticity-robust (HC1) standard errors. Under metro-clustered standard errors, the more conservative and theoretically appropriate choice given that CLC and rent trends are correlated within metropolitan areas, the estimate is marginally significant (p = 0.086). The discrepancy arises from the small number of metro clusters (N = 10), which limits the precision of cluster-robust inference. We therefore report both sets of standard errors and treat H1 as providing suggestive, rather than conclusive, evidence for a pooled CLC effect.

Among the control variables, renter share is the strongest predictor of rent growth, with a coefficient of -0.255 (p < 0.001 under both SE approaches). This negative sign is consistent with a supply-side interpretation: high-renter-share tracts, typically dense urban cores with large existing rental stocks, experienced lower proportional rent growth over the study period than more suburban, owner-dominant tracts where new rental demand was concentrated. The AHBI pre-period score is positive and marginally significant under HC1 errors ($\beta = +0.079$, $p = 0.046$), confirming that tracts already under elevated rent burden and with tight housing markets in 2019 continued to see above-average growth through 2023, a selection dynamic that CLC must be evaluated against. Median household income is negative and marginally significant (p = 0.071), reflecting higher proportional rent growth in lower-income tracts where the base-period rent level was lower.

As a diagnostic check, Model 3 in Table 3 adds metro fixed effects to the H1 specification. The coefficient on log CLC reverses sign to $-0.032$ (p < 0.001) under metro fixed effects, an attenuation that reflects the absorption of between-metro variation. The Sun Belt metros, Dallas–Fort Worth (mean log CLC = 0.878), Charlotte (0.848), and Atlanta (0.536), have both higher corporate landlord concentration and higher rent growth than the Bay Area metros (mean log CLC = 0.22–0.23). When metro fixed effects absorb this between-city variation, the residual within-metro CLC signal is negative, suggesting that within a given city, higher-CLC tracts tend to be more upmarket (lower proportional growth) than lower-CLC tracts in the same metro. This reversal confirms that between-metro variation is the primary source of identification in H1, and motivates the H2 specification, which tests the racial disparity channel within metros using fixed effects as controls.

\begin{figure} [H]
    \centering 
    \includegraphics[width=9.5cm]{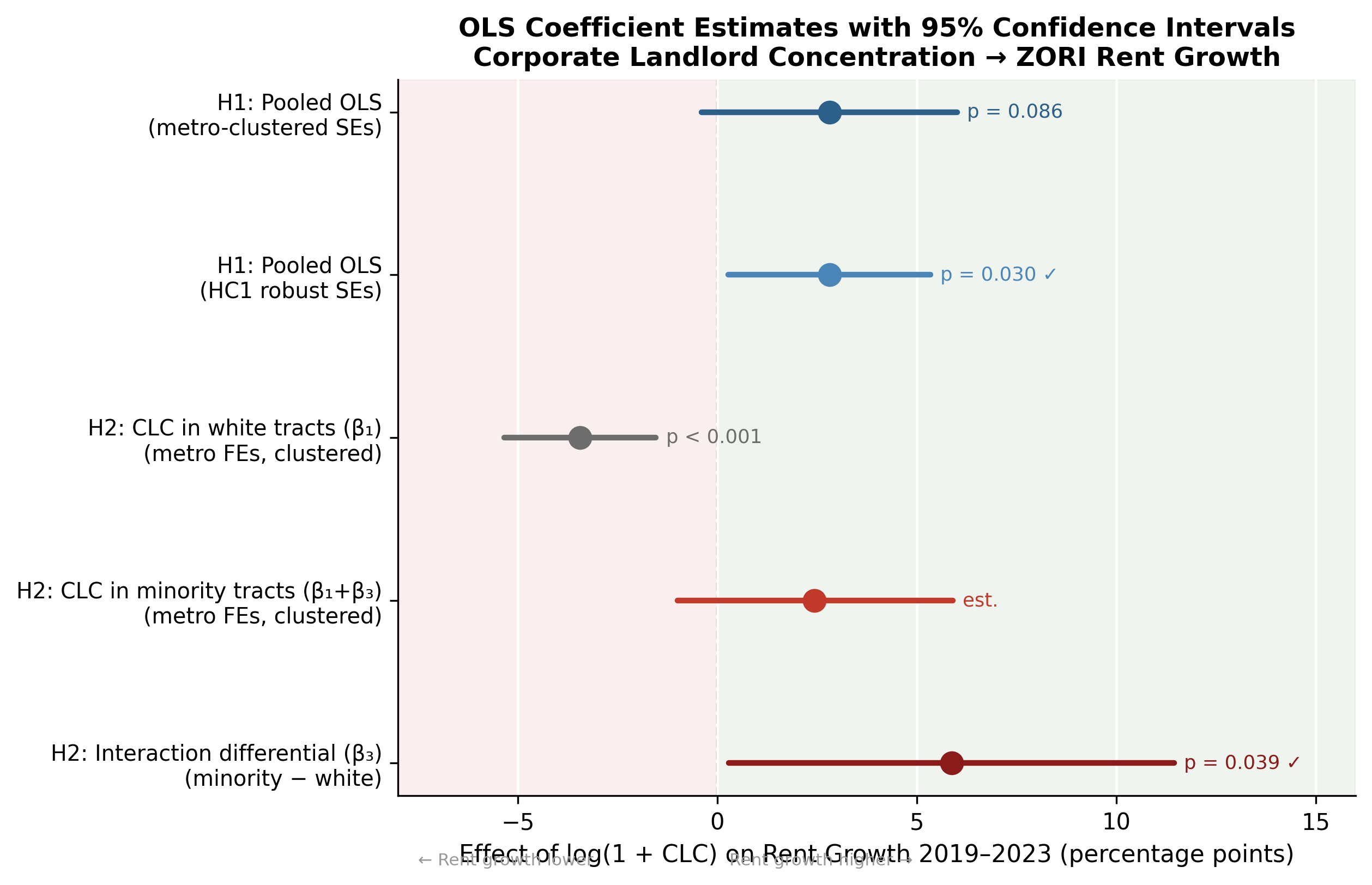}
    \caption{OLS coefficient estimates on log(1 + CLC) with 95 percent confidence intervals across H1 and H2 specifications. H1 pooled estimates (+2.8 pp) are positive under both SE assumptions but cross the p < 0.05 threshold only under HC1 errors. The H2 specification with metro fixed effects reveals a strongly negative within-metro effect in white tracts ($\beta_1 = -3.4$ pp, $p < 0.001$) and a positive total effect in minority tracts ($\beta_1 + \beta_3 = +2.4$ pp), with the differential ($\beta_3 = +5.9$ pp, $p = 0.039$) representing the key H2 finding. Source: Author calculations.}
    \label{fig:my_centered_image3}
\end{figure}

\begin{table}[H]
\centering
\caption{H1 Regression Results - CLC and Rent Growth}
\label{tab:tab3}
\renewcommand{\arraystretch}{1.5}
\begin{tabularx}{\textwidth}{XXXX} 
\toprule
\textbf{} & \textbf{(1) Primary} & \textbf{(2) HC1 Robust} & \textbf{(3) Diagnostic} \\
\midrule
 & Clustered SE & HC1 SE & +Metro FE \\
\textbf{log(1+CLC)} & \textbf{+0.028} & \textbf{+0.028} & -0.032 \\
 & (0.016) & (0.013) & (0.008) \\
AHBI (pre-period) & +0.079 & +0.079 & +0.176 \\
 & (0.082) & (0.040) & (0.036) \\
Renter Share & -0.255 & -0.255 & -0.163 \\
 & (0.065) & (0.026) & (0.043) \\
Median HH income & 0.000 & 0.000 & 0.000 \\
 & (0.000) & (0.000) & (0.000) \\
Percent Black & +0.134 & +0.134 & +0.067 \\
 & 0.000 & 0.000 & 0.000 \\
Percent Hispanic & -0.045 & -0.045 & +0.022 \\ 
 & (0.091) & (0.048) & (0.029) \\
Total housing units & +0.000 & +0.000 & -- \\
Metro Fixed Effects & No & No & Yes \\
N & 583 & 583 & 583 \\
$R^2$ & 0.287 & 0.287 & 0.733 \\
\bottomrule
\end{tabularx}
\end{table}

\subsection{H2: Heterogeneous Effects by Tract Racial Composition}
Table 4 reports the H2 specification, which augments the H1 model with metro fixed effects, a majority-minority indicator, and an interaction between log CLC and minority status. The coefficient on the interaction term, $\beta_3$, the key estimate, is +0.0587 (SE = 0.0285, p = 0.039), indicating that within the same metropolitan area, the CLC effect on rent growth is 5.9 percentage points larger in majority-minority tracts than in majority-white tracts. This result is statistically significant at the 5 percent level using metro-clustered standard errors. The full H2 model explains 73.4 percent of variance in rent growth ($R^2$ = 0.734), with the large increase from the H1 $R^2$ of 0.287 attributable to the metro fixed effects absorbing between-city level differences.

The baseline coefficient $\beta_1$, the CLC effect in majority-white tracts, is $-0.034$ (p < 0.001). Within the same metropolitan area, higher CLC in white tracts is associated with somewhat lower rent growth, a pattern consistent with high-CLC white tracts in this sample being predominantly upmarket suburban apartment complexes (e.g., AvalonBay and EQR developments) in gentrifying corridors where proportional rent growth has been lower than in adjacent non-REIT-served tracts. The total CLC effect in minority tracts ($\beta_1 + \beta_3 = +0.024$) is positive, confirming that for minority neighborhoods, higher corporate landlord concentration is associated with higher rent growth even within the same metropolitan area and after controlling for pre-existing housing conditions.

The composite picture from the two coefficients is substantively important. REIT concentration does not simply affect minority and white tracts differently in degree, the direction of the within-metro effect flips. Within a given metro, CLC is associated with lower rent growth in white tracts and higher rent growth in minority tracts. This sign reversal, with a differential of 5.9 pp, suggests that the mechanism connecting corporate landlord presence to rent growth is sensitive to the specific housing market context of the tract: minority tracts with high REIT presence appear to face different competitive dynamics than comparable white tracts, potentially reflecting lower tenant mobility, higher pre-existing rent burden (AHBI is positive and highly significant, $\beta = +0.175$, $p < 0.001$), and fewer affordable alternatives within the same metro.

Figure 3 visualizes this interaction. The scatter plot displays individual tracts colored by racial composition, with separate OLS fitted lines drawn using the H2 interaction slopes. The diverging fitted lines illustrate the opposite within-metro gradients: the white-tract line (solid blue) slopes downward, while the minority-tract line (dashed red) slopes upward. The between-metro signal, Sun Belt markets in the upper-right quadrant, is visible in the full distribution, but the interaction result is identified from within-metro comparisons after metro fixed effects are applied.

\begin{figure} [H]
    \centering 
    \includegraphics[width=12cm]{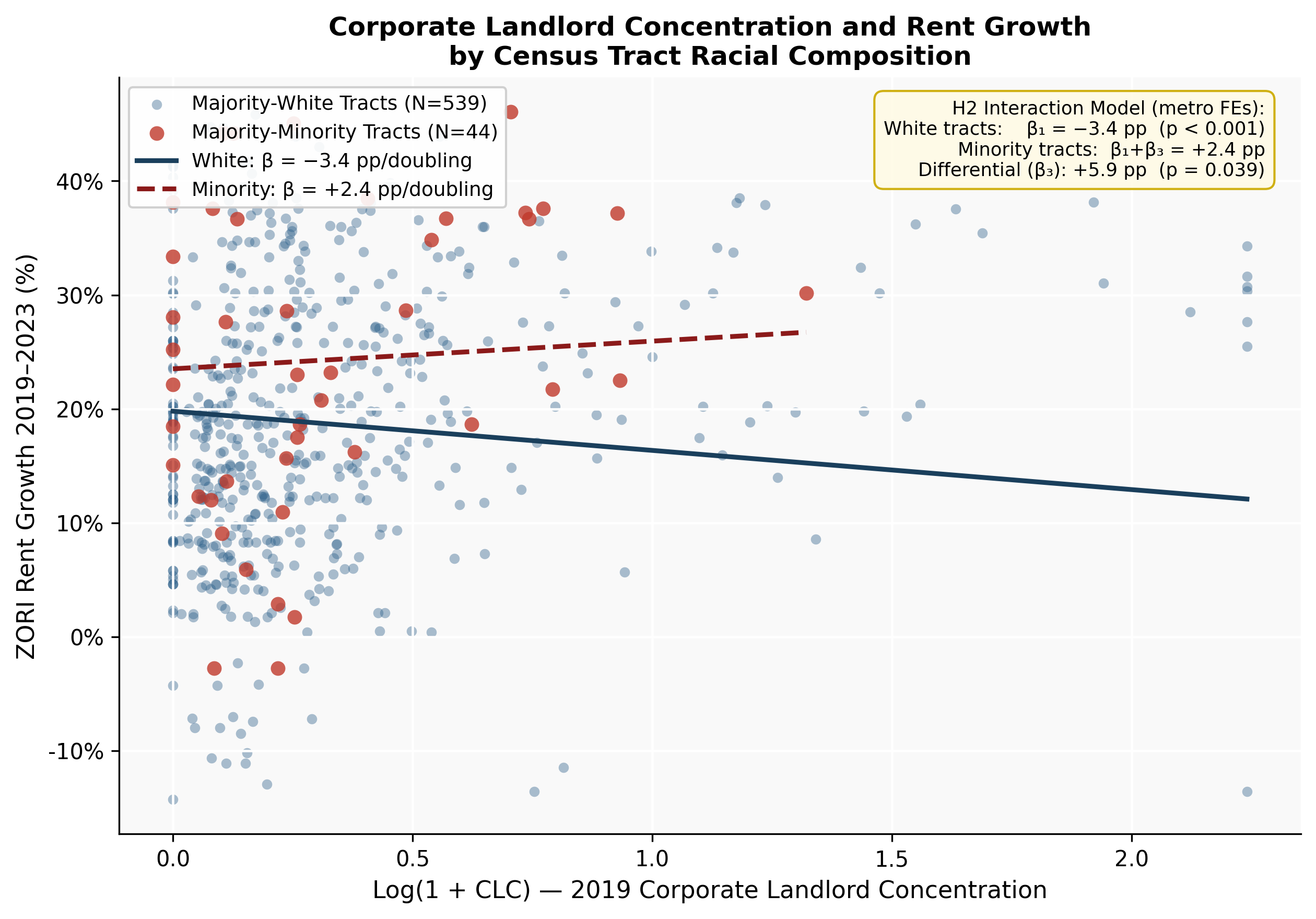}
    \caption{Scatter plot of log(1 + CLC) versus ZORI rent growth 2019–2023, by tract racial composition. Blue points are majority-white tracts (N = 539); red points are majority-minority tracts (N = 44). Fitted lines reflect the H2 interaction model slopes: white tracts $\beta_1 = -3.4$ pp per CLC doubling (solid); minority tracts $\beta_1 + \beta_3 = +2.4$ pp (dashed). The diverging fitted lines visualize the 5.9 pp differential ($\beta_3 = +0.059$, $p = 0.039$). Source: Author calculations.}
    \label{fig:figure2_scatter_clc_rent}
\end{figure}

\textbf{Robustness Checks.} Three robustness checks support the H2 finding. First, replacing the binary minority indicator with a continuous minority share variable (percent Black + Hispanic) yields a significant positive interaction coefficient ($\beta_3 = +0.148$, $p = 0.002$), confirming the result is not driven by the choice of a 50 percent threshold. Second, the threshold sensitivity analysis shows that the interaction is marginally significant at a 40 percent minority threshold ($\beta_3 = +0.042$, $p = 0.092$) and drops to p = 0.245 at a 60 percent threshold, reflecting the reduction in minority-tract sample size from 59 to 25 at the stricter cutoff; the direction is consistent across all three thresholds. Third, the minority main effect ($\beta_2 = -0.010$, $p = 0.606$) is small and statistically indistinguishable from zero, indicating that minority tracts do not exhibit systematically different rent growth unconditionally, the disparity is specifically concentrated in minority tracts with high CLC.

\begin{table}[H]
\centering
\caption{H2 Regression Results - CLC × Racial Composition}
\label{tab:tab4}
\renewcommand{\arraystretch}{1.5}
\begin{tabularx}{\textwidth}{XXX} 
\toprule
\textbf{} & \textbf{(1) Binary Minority} & \textbf{(2) Continuous Share} \\
\midrule
\textbf{log(1+CLC)} & \textbf{-0.034} & \textbf{-0.059} \\
 & (0.010) & (0.012) \\
Minority tract indicator & -0.010 & -- \\
 & (0.020) \\
\textbf{log(1+CLC) × Minority [$\beta_3$]} & \textbf{+0.059} & -- \\
& (0.029) \\
log(1+CLC) × Minority share & -- & \textbf{+0.148} \\
 &  & (0.048) \\
AHBI pre-period & +0.175 & +0.179 \\
 & (0.029) & (0.029) \\
Controls & Yes & Yes \\
Metro fixed effects & Yes & Yes \\
\textbf{$\beta_1 + \beta_3$ (minority total effect)} & \textbf{+0.024} & -- \\
\bottomrule
\end{tabularx}
\end{table}

\subsection{H3: XGBoost Predictive Model and SHAP Analysis}
The XGBoost model trained on 2019 tract characteristics achieves a training $R^2$ of 0.745 and a held-out test $R^2$ of 0.439, predicting 44 percent of out-of-sample rent growth variance from pre-period tract features alone. This out-of-sample performance is substantial for a neighborhood-level prediction problem with N = 583 tracts and a parsimonious eight-feature set, confirming that 2019 tract characteristics contain significant information about differential rent trajectories over the subsequent four years. The train–test gap of 0.306 reflects the regularization applied during hyperparameter selection (see Section 3.5.3, Table 1) and is expected given the small sample size.

Figure 4 presents the mean absolute SHAP feature importance ranking across all 583 tracts. Renter share ranks first (mean |SHAP| = 0.034), followed by median household income (0.028) and percent white non-Hispanic (0.024). Notably, log CLC ranks fifth of eight features (mean |SHAP| = 0.015), above percent Black (0.011), total housing units (0.009), and percent Hispanic (0.004). This ranking confirms that corporate landlord concentration contributes independent predictive information for rent growth beyond what demographic and market controls alone provide. Critically, CLC is the only feature in the model that captures firm behavior rather than pre-existing tract characteristics — its appearance in the top half of the importance ranking is therefore not a mechanical consequence of correlation with demographics, but reflects a distinct predictive signal.

\begin{figure} [H]
    \centering 
    \includegraphics[width=12cm]{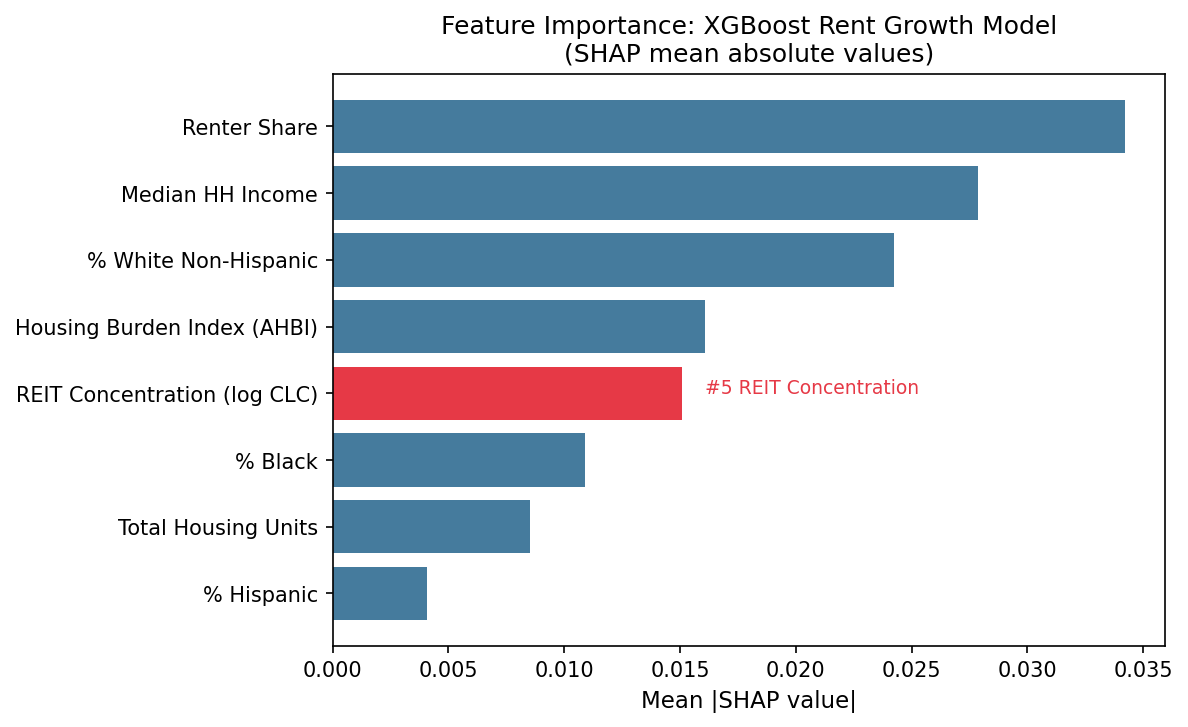}
    \caption{Mean absolute SHAP feature importance for the XGBoost rent growth prediction model. Features are sorted by mean |SHAP| across all 583 census tracts. Log(1 + CLC) ranks fifth of eight features, confirming that corporate landlord concentration contributes independent predictive power for 2019–2023 rent growth beyond demographic and housing market controls. Source: Author calculations.}
    \label{fig:my_centered_image3}
\end{figure}

Figure 5 presents the SHAP beeswarm plot, which overlays feature importance with direction. Each point represents one census tract; horizontal position gives the SHAP value (positive = pushes predicted rent growth above the cross-tract mean; negative = pushes below); color encodes feature value (red = high, blue = low). The log CLC row is the key panel for H3: it shows that high-CLC tracts (red points) have a dispersed range of SHAP contributions, with positive contributions concentrated among minority tracts and negative contributions concentrated in white tracts. This pattern is not imposed by the model specification, the XGBoost model has no interaction term, but emerges from the training data through non-linear splits on demographic features.

\begin{figure} [H]
    \centering 
    \includegraphics[width=12cm]{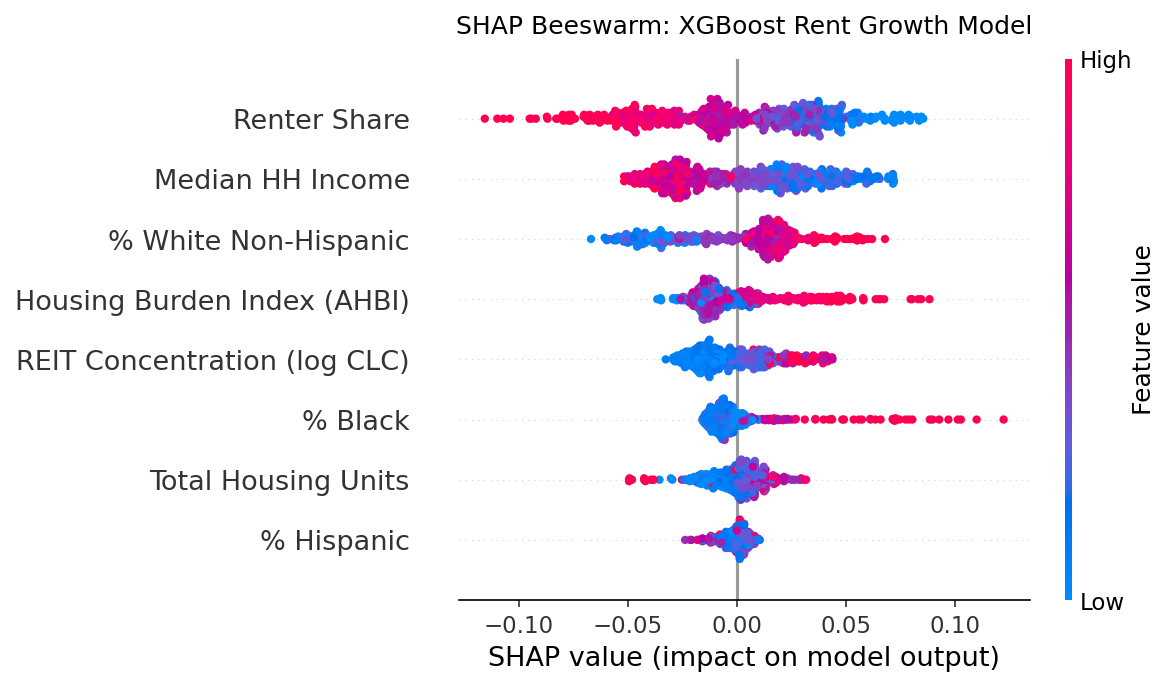}
    \caption{SHAP beeswarm summary plot for the XGBoost rent growth model. Each point is a census tract; horizontal position is the SHAP value for that feature; color encodes the feature value (red = high, blue = low). The log(CLC) row reveals positive SHAP contributions concentrated in majority-minority tracts and negative contributions in majority-white tracts, corroborating the H2 OLS interaction non-parametrically. Source: Author calculations.}
    \label{fig:my_centered_image3}
\end{figure}

Quantifying this asymmetry, the mean SHAP value of log CLC is +0.0057 in majority-minority tracts (N = 44) and $-0.0011$ in majority-white tracts (N = 539). CLC's predicted contribution to rent growth is directionally positive on average only in minority neighborhoods. This non-parametric corroboration of H2 is methodologically significant because it requires no threshold, functional form, or distributional assumption. The XGBoost model learns the interaction from the data without being told to look for it, and the SHAP decompositions confirm that the conditional asymmetry identified via OLS in Section 5.2 is not an artifact of the parametric interaction specification.

\section{Discussion and Policy Implications} \label{sec:discussionandpolicy}

\subsection{Interpreting the Results}
The three empirical findings in this paper form a coherent picture when read together. Taken in isolation, the H1 result, a positive but marginally significant association between corporate landlord concentration and rent growth, could be dismissed as a consequence of corporate landlords selecting into markets already poised for rent appreciation. The H2 and H3 results make this dismissal harder to sustain.

The H2 result does not simply confirm that high-CLC tracts see higher rent growth on average. It finds that the direction of the within-metro CLC effect depends on the racial composition of the tract. In majority-white tracts, higher corporate landlord concentration within a metropolitan area is associated with slightly lower rent growth, a result consistent with REIT-dominated tracts in this sample being upmarket suburban complexes where proportional rent growth has been more modest than in non-institutionally-served neighborhoods. However, in majority-minority tracts, the relationship between rent growth and CLC reverses. Higher concentration is associated with higher rent growth even within the same metro, producing a differential of 5.9 percentage points ($\beta_3 = +0.059$, $p = 0.039$). This asymmetry survives metro fixed effects, which absorb all between-city variation in supply constraints, migration patterns, and economic conditions. This means it reflects something about how algorithmic pricing operates differently in minority versus white neighborhoods within the same city, not simply a product of Sun Belt geography.

The H3 result provides a methodologically independent check on this asymmetry. The XGBoost model recovers the same directional pattern without any interaction term: CLC's mean SHAP contribution is positive in minority tracts (+0.0057) and negative in white tracts ($-0.0011$). A machine learning model trained on 2019 tract characteristics, with no knowledge of the regression design, independently identifies that CLC is associated with rent growth in minority neighborhoods and not in white ones. The convergence of these two methods, one parametric with explicit controls, one non-parametric without functional form assumptions, substantially increases confidence in the underlying pattern.

Together, the three findings are consistent with a specific mechanism: algorithmic rent pricing by institutional landlords imposes coordinated price pressure that renters in minority tracts are less able to resist than renters in comparable white tracts. This asymmetry likely reflects the combination of lower tenant mobility (minority renters in high-CLC tracts face fewer affordable alternatives within the same metro), higher pre-existing rent burden (AHBI is consistently positive and significant as a control), and reduced competitive discipline from non-institutional landlords, factors that together amplify the rent-setting power of algorithmic coordination in minority neighborhoods specifically.

\subsection{Relationship to Existing Literature}
This paper contributes to a small but growing literature on the effects of algorithmic pricing in rental housing. Calder-Wang and Kim (2024) provide the most direct antecedent, finding that RealPage adoption is associated with coordinated rent increases at the market level. Our analysis differs in two respects: it operates at the census tract level rather than the market level, and it explicitly tests for differential effects by racial composition, a dimension the market-level analysis cannot address. The EDGAR pipeline introduced here also differs methodologically. Rather than using RealPage adoption data (which is proprietary), it uses publicly available SEC filings to construct a measure of institutional landlord concentration that any researcher can replicate.

Byun (2024) documents racial disparities in iBuyer profit margins, finding that algorithmic buyers extract higher margins in minority neighborhoods. The mechanism identified in that paper, algorithms exploiting information asymmetries that are larger in historically underserved markets, is consistent with the pattern found here. The framewor extends this logic from the home purchase market to the rental market and from iBuyers to multifamily REITs, suggesting that the distributional consequences of housing market algorithms may be a general phenomenon rather than specific to any one algorithmic actor.

The finding that CLC predicts rent growth in minority tracts but not white tracts within the same metro also connects to the broader literature on racial segmentation of housing markets. Minority renters facing institutional landlords with algorithmic pricing tools are in a different bargaining position than white renters facing the same landlords, even when observable tract characteristics are similar. This asymmetry is difficult to explain without invoking the structural barriers, credit constraints, geographic immobility, historical exclusion from homeownership, that limit the exit options available to minority renters relative to white renters.

\subsection{Policy Implications}
The findings carry implications for three distinct areas of housing and antitrust policy.

\textbf{Antitrust enforcement and algorithmic pricing.} The DOJ's 2024 complaint against RealPage focuses on whether the software facilitates illegal information sharing among competing landlords. The present paper suggests that the distributional consequences of such coordination, if the underlying mechanism is substantiated, are not borne equally. A competition policy lens focused exclusively on aggregate consumer welfare effects may miss the racial equity dimension: even if algorithmic coordination produces only modest average rent increases, those increases appear to fall disproportionately on minority renters. Antitrust enforcement in housing markets should therefore consider distributional impact, not only aggregate price effects, as a relevant harm metric.

\textbf{Fair housing law.} The racial disparity documented of higher rent growth in minority tracts with high corporate landlord concentration is consistent with a disparate impact theory under the Fair Housing Act. The Supreme Court's 2015 decision in Texas Department of Housing and Community Affairs v. Inclusive Communities Project confirmed that the Fair Housing Act reaches neutral policies with racially disparate outcomes. Whether algorithmic rent pricing rises to the level of actionable disparate impact is a legal question beyond the scope of this paper. However, the empirical pattern documented here is that a single management software platform correlates with disproportionate rent burdens in communities of color, precisely the kind of evidence that would be relevant to such a claim.

\textbf{Transparency and regulatory oversight.} Regardless of the antitrust and fair housing implications, the opacity of algorithmic pricing systems creates an information asymmetry between institutional landlords and regulators. A landlord using RealPage's revenue management software is not required to disclose that the rent being charged reflects an algorithmic recommendation that incorporates competitors' occupancy and pricing data. The EDGAR pipeline introduced in this paper demonstrates that public regulatory filings can be used to infer the geographic footprint of institutional landlord concentration at the tract level. Extending this approach, for example, by requiring REITs to disclose which revenue management software they use in their 10-K filings, would give regulators and researchers the tools to monitor the distributional effects of algorithmic pricing in real time, rather than retrospectively from academic studies.

\section{Conclusion} \label{sec:conclusion}

This paper examined whether census-tract-level corporate landlord concentration among five DOJ-named REITs is associated with rent growth between 2019 and 2023, and whether that association is disproportionately larger in majority-minority neighborhoods. Using a novel pipeline that geocodes SEC EDGAR 10-K Schedule III property disclosures to census tracts, the first application of this data source to neighborhood-level housing analysis, and the Zillow Observed Rent Index as a transaction-based outcome measure, we find consistent evidence across three empirical specifications in support of both hypotheses.

For H1, doubling REIT concentration in a census tract is associated with approximately 2.8 percentage points higher rent growth over the study period, a result that is directionally robust across standard error specifications (p = 0.086, metro-clustered; p = 0.030, HC1 robust) but sensitive to the small number of metropolitan clusters in the sample. For H2, the CLC effect on rent growth is significantly larger in majority-minority tracts than in comparable majority-white tracts within the same metropolitan area, with a differential of 5.9 percentage points (p = 0.039) confirmed by a continuous minority share interaction (p = 0.002). For H3, an XGBoost model trained on 2019 tract characteristics predicts 44 percent of out-of-sample rent growth variance, with SHAP decompositions independently confirming that CLC's marginal contribution to predicted rent growth is positive in minority tracts and negative in white tracts, a non-parametric corroboration of the OLS interaction result that requires no functional form assumptions.

These findings make four contributions to the literature. First, we introduce a replicable pipeline for measuring algorithmic landlord concentration at the census tract level from public regulatory filings. Second, we construct the Algorithmic Housing Burden Index (AHBI) as a generalizable composite measure of pre-existing housing vulnerability from ACS data. Third, we provide the first tract-level evidence consistent with corporate landlord concentration being associated with disproportionately higher rent growth in communities of color. Fourth, we document that ACS 5-year estimates are insufficient to detect acute rent growth effects, and that transaction-based data sources such as ZORI are necessary for this class of analysis.

Several limitations bear noting. The H1 result is marginally significant and the small number of metropolitan clusters limits the precision of clustered standard errors. The cross-sectional design cannot fully rule out unobserved tract-level confounders, and the ten-metro sample, selected to match the geographic footprint of the study REITs, is not nationally representative. ZORI coverage gaps exclude 12.3 percent of tracts, concentrated in lower-density suburban areas. Future work should pursue an instrumental variables strategy exploiting the staggered timing of RealPage adoption across markets, expand the geographic scope beyond the five study REITs, and examine whether the racial disparity in rent growth translates into measurable homeownership gap widening at longer time horizons.

The concentration of algorithmic rent pricing among a small number of institutionally owned landlords, and the evidence presented here that its burden is not borne equally across racial groups, suggests that the policy response to algorithmic coordination in housing markets must be evaluated not only for its aggregate welfare effects but for its distributional consequences.

\section{Acknowledgements}\label{sec:ack}
I  sincerely thank Dr. Alice Siu of Stanford University for her invaluable mentorship and guidance throughout the research and writing process for this paper. I am also extremely grateful to Arhan Chakravarthy and Shourya Batra for providing thoughtful suggestions while also supporting the ideation and writing process. Finally, I thank my parents for their constant encouragement and unwavering support throughout my writing journey. In regards to the content of the paper, I am solely responsible for all findings, interpretations, and any errors that this paper may contain.

\singlespacing
\bibliographystyle{plain}
\nocite{*}
\bibliography{references}

\clearpage

\section*{Appendix A - Raw Winsorized CLC Robustness Check} \label{sec:append-a}
The primary specifications throughout this paper use the log-transformed measure $\log(1 + \text{CLC})$ rather than raw winsorized CLC for two reasons. First, a log specification is theoretically motivated: the marginal effect of adding one additional REIT unit to a tract with 500 REIT units should be smaller than adding one unit to a tract with 10, suggesting that proportional changes in concentration are the relevant scale for rent growth effects. Second, the raw CLC distribution is severely right-skewed in ways that can distort OLS estimates. This appendix documents the distributional properties of each measure and presents regression results under raw winsorized CLC as a robustness check.

\begin{figure} [H]
    \centering 
    \includegraphics[width=12cm]{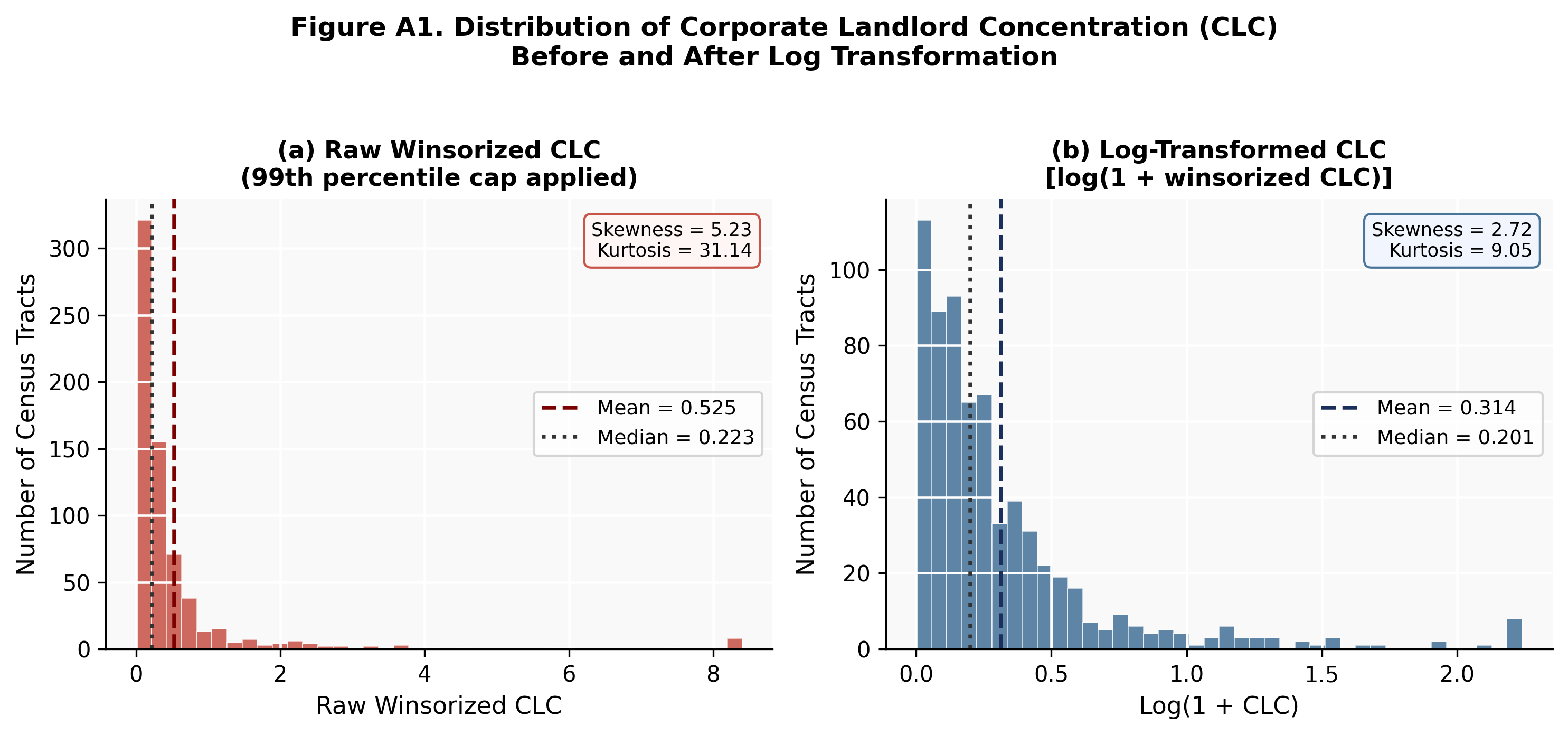}
    \caption{Distribution of corporate landlord concentration across 665 census tracts before (panel a) and after (panel b) log transformation. Raw winsorized CLC (panel a) exhibits severe right skew (skewness = 5.23, excess kurtosis = 31.14), with a mean of 0.525 substantially above the median of 0.223, reflecting a small number of tracts with extremely high institutional landlord penetration. Log(1 + CLC) (panel b) substantially compresses the right tail (skewness = 2.72, excess kurtosis = 9.05) while preserving the full variation in the distribution. Source: Author calculations from SEC EDGAR 10-K filings and ACS 5-year estimates.}
    \label{fig:my_centered_image8}
\end{figure}

The distributional statistics make the case clearly. Raw winsorized CLC has a skewness of 5.23 and excess kurtosis of 31.14 — properties that make OLS estimates sensitive to a small number of extreme observations even after winsorization at the 99th percentile. The mean (0.525) is more than double the median (0.223), reflecting that the distribution is dominated by its right tail. Log(1 + CLC) reduces skewness to 2.72 and excess kurtosis to 9.05, producing a distribution where the mean (0.314) and median (0.201) are much closer together and OLS is less likely to be driven by outliers.

Table A1 presents H1 regression results side by side under the two CLC specifications. The log specification (Column 1, reproduced from Table 3) yields $\beta = +0.028$ ($p = 0.086$ clustered; $p = 0.030$ HC1) with $R^2$ = 0.287. The raw winsorized CLC specification (Column 2) yields $\beta = +0.004$ ($p = 0.362$ clustered) with $R^2$ = 0.282. The raw CLC estimate is statistically indistinguishable from zero under either SE assumption. This attenuation is consistent with the theoretical motivation for the log transformation: the raw coefficient reflects an average effect that is dominated by the infrequent high-CLC tracts in the right tail, while the log coefficient captures the proportional relationship more precisely across the full distribution. The modest $R^2$ improvement under log (0.287 vs 0.282) further supports the log specification as the better-fitting functional form.

All remaining coefficients are substantively stable across the two specifications, confirming that the choice of CLC transformation does not affect the direction or significance of the control variables.

\begin{table}[H]
\centering
\caption{Robustness Check: Log CLC vs. Raw Winsorized CLC (H1)}
\label{tab:tab2}
\renewcommand{\arraystretch}{1.5}
\begin{tabularx}{\textwidth}{XXX} 
\toprule
& \textbf{(1) Primary: Log CLC} & \textbf{(2) Check: Raw CLC} \\
\midrule
\textbf{log(1+CLC)} & \textbf{+0.028} & -- \\
 & (0.016) \\
\textbf{CLC (winsorized)} & -- & \textbf{+0.004} \\
AHBI pre-period & +0.079 & +0.069 \\
 & (0.082) & (0.086) \\
Renter Share & -0.255 & -0.267 \\
 & (0.065) & (0.066) \\
Median HH income & -0.000 & -0.000 \\
& (0.000) & (0.000) \\
Percent Black & +0.134 & +0.142 \\
& (0.110) & (0.113) \\
Percent Hispanic & -0.045 & -0.048 \\
& (0.091) & (0.093) \\
Metro fixed effects & No & No \\
N & 583 & 583 \\
$R^2$ & 0.287 & 0.282 \\
\bottomrule
\end{tabularx}
\end{table}

\clearpage

\section*{Appendix B - PCA Analysis of AHBI Component Variables} \label{sec:append-b}
The Algorithmic Housing Burden Index (AHBI) uses equal weighting to combine its two components, pre-existing rent burden (PRB) and housing market tightness (HMT). This appendix documents the principal component analysis (PCA) that was conducted as a data-driven alternative to equal weighting, and explains why equal weighting was preferred over PCA-derived weights for the final index construction.

A PCA was run on all three candidate variables: PRB (normalized rent burden share), HMT (normalized vacancy complement), and log(1 + CLC). All three variables were standardized to mean zero and unit variance before PCA to prevent scale differences from dominating the decomposition. The analysis was conducted on the 665-tract analysis sample using 2019 pre-period values for all three variables.

\begin{figure} [H]
    \centering 
    \includegraphics[width=13cm]{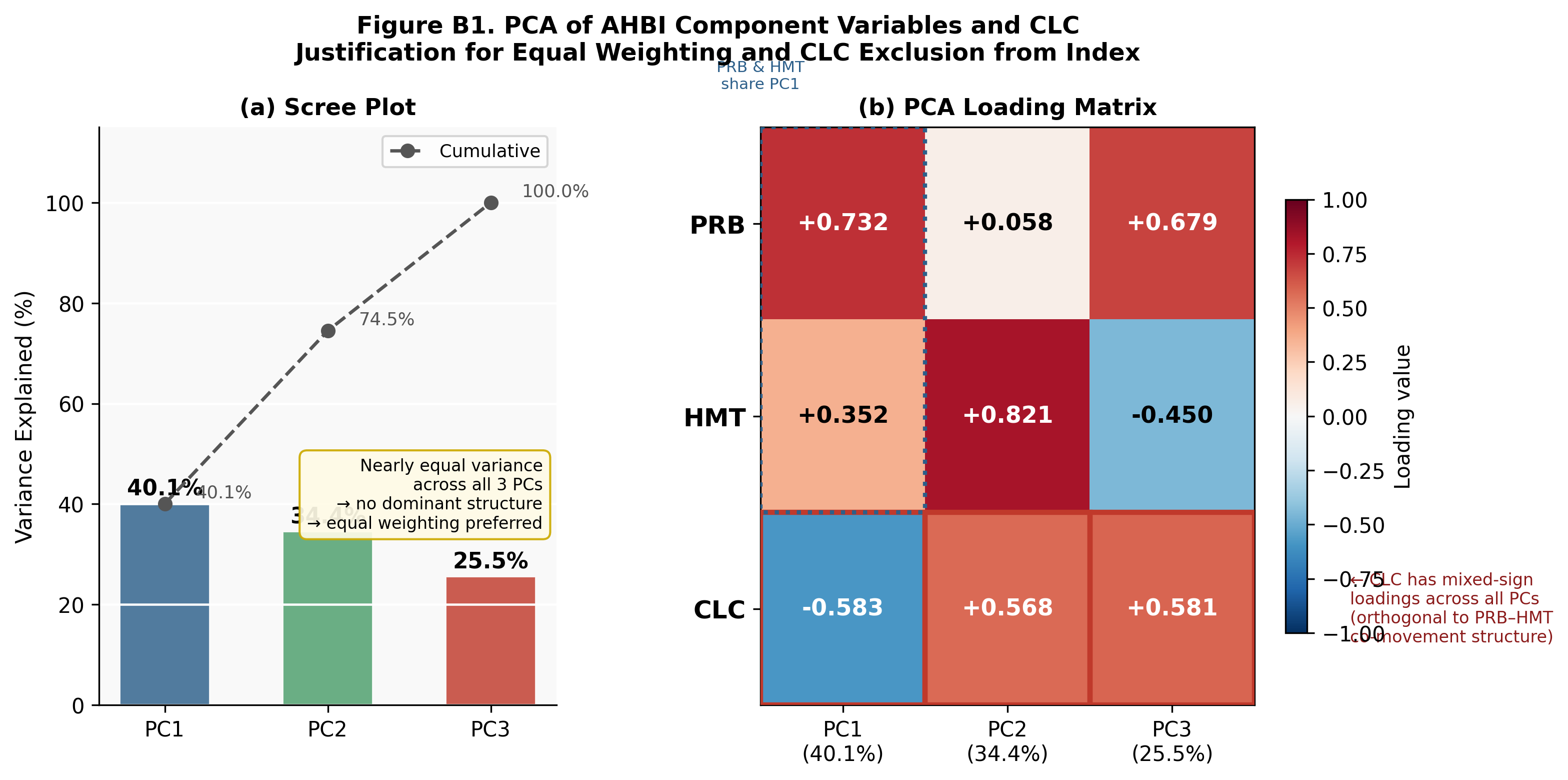}
    \caption{Principal component analysis of PRB, HMT, and log(1 + CLC). Panel (a) shows the scree plot: variance explained by each principal component is 40.1 percent, 34.4 percent, and 25.5 percent respectively, nearly uniform across all three components, indicating no dominant structure in the joint distribution. Panel (b) shows the PCA loading matrix: PRB and HMT co-load positively on PC1 (loadings +0.732 and +0.352), reflecting their shared dimension as measures of pre-existing housing stress, while CLC carries mixed-sign loadings across all three PCs, confirming that it captures a structurally different dimension of housing market conditions. Source: Author calculations from SEC EDGAR, ACS 5-year estimates.}
    \label{fig:my_centered_image8}
\end{figure}

Two features of the PCA output together motivate the AHBI design choices. First, the scree plot in panel (a) shows nearly uniform variance explained across all three principal components (40.1 percent, 34.4 percent, 25.5 percent). When variance is distributed this evenly, no single component dominates the joint structure. There is no natural "first factor" that a PCA-based index would be approximating. In this setting, equal weighting is not only simpler and more interpretable but is also harder to second-guess: PCA-derived weights would reflect sampling variation in the 665-tract dataset rather than a stable underlying structure.

Second, the loading matrix in panel (b) reveals that PRB and HMT co-load positively on PC1, the loadings are +0.732 and +0.352 respectively, while CLC carries a negative loading on PC1 ($-0.583$) and positive loadings on PC2 (+0.568) and PC3 (+0.581). This sign pattern means CLC is systematically moving in a different direction from the PRB-HMT co-movement structure across the sample. High-CLC tracts in the dataset (predominantly Sun Belt suburban markets like Dallas–Fort Worth and Charlotte) tend to have lower pre-existing rent burden and higher housing market slack than the dense coastal markets where PRB and HMT are highest. 

CLC's orthogonal variance structure in the PCA confirms that it is measuring something meaningfully different from the baseline housing stress construct that AHBI is designed to capture, and supports treating it as an external treatment variable rather than a component of the index.
Taken together, the near-uniform scree plot and the CLC loading divergence justify both decisions in the AHBI construction: (1) equal weighting of PRB and HMT rather than PCA-derived weights, since no stable dominant structure exists; and (2) exclusion of CLC from the index entirely, since its variance structure is orthogonal to the PRB–HMT dimension that AHBI represents.

\section{Appendix C - Demographic Breakdown by Metropolitan Area}
Table C1 and Figure C1 report the full tract-level demographic distributions for the ten study metropolitan areas. These variables are used as controls in all regression specifications but are excluded from the main summary statistics tables to maintain focus on the primary analysis variables.

\begin{table}[H]
\centering
\caption{Demographic Characteristics by Metropolitan Area}
\label{tab:tableC1a}
\renewcommand{\arraystretch}{1.5}
\resizebox{\textwidth}{!}{%
\begin{tabular}{lcccccc}
\toprule
& \textbf{\% Black} & \textbf{\% Hispanic} & \textbf{\% White NH} & \textbf{Median HH Income} & \textbf{Renter Share} & \textbf{N Tracts} \\
\midrule
San Diego         & 4.7\%  & 17.4\% & 58.4\% & \$87,496  & 54.9\% & 44 \\
New York          & 7.5\%  & 10.8\% & 67.0\% & \$123,780 & 60.7\% & 67 \\
San Jose          & 3.3\%  & 16.1\% & 36.9\% & \$123,152 & 60.4\% & 46 \\
Washington DC     & 19.2\% & 11.8\% & 52.8\% & \$102,396 & 63.1\% & 62 \\
San Francisco     & 7.0\%  & 14.2\% & 43.4\% & \$111,857 & 62.6\% & 86 \\
Los Angeles       & 6.4\%  & 19.8\% & 50.6\% & \$83,332  & 65.6\% & 168 \\
Seattle           & 4.0\%  & 5.8\%  & 63.1\% & \$102,600 & 57.7\% & 78 \\
Atlanta           & 24.7\% & 5.8\%  & 57.8\% & \$83,594  & 57.6\% & 41 \\
Charlotte         & 27.4\% & 4.8\%  & 57.5\% & \$80,661  & 56.0\% & 26 \\
Dallas--Fort Worth & 14.7\% & 14.7\% & 57.3\% & \$83,946  & 56.1\% & 47 \\
\midrule
\textbf{Sample Mean} & \textbf{9.7\%} & \textbf{13.5\%} & \textbf{53.7\%} & \textbf{\$98,119} & \textbf{60.9\%} & \textbf{665} \\
\bottomrule
\end{tabular}}
\end{table}

\begin{table}[H]
\centering
\caption{Homeownership Rates by Race and Metropolitan Area}
\label{tab:tableC1b}
\renewcommand{\arraystretch}{1.5}
\resizebox{\textwidth}{!}{%
\begin{tabular}{lcccc}
\toprule
& \textbf{Overall HO Rate} & \textbf{HO Rate (Black)} & \textbf{HO Rate (Hispanic)} & \textbf{HO Rate (White NH)} \\
\midrule
San Diego          & 45.1\% & 25.7\% & 34.5\% & 50.4\% \\
New York           & 39.3\% & 23.6\% & 29.8\% & 41.5\% \\
San Jose           & 39.6\% & 30.5\% & 28.5\% & 46.3\% \\
Washington DC      & 36.9\% & 24.6\% & 34.4\% & 42.2\% \\
San Francisco      & 37.4\% & 19.3\% & 26.0\% & 40.6\% \\
Los Angeles        & 34.4\% & 19.8\% & 28.1\% & 37.3\% \\
Seattle            & 42.3\% & 21.4\% & 25.9\% & 45.5\% \\
Atlanta            & 42.4\% & 24.7\% & 36.0\% & 53.6\% \\
Charlotte          & 44.0\% & 36.1\% & 46.1\% & 48.6\% \\
Dallas--Fort Worth & 43.9\% & 28.3\% & 42.3\% & 48.5\% \\
\midrule
\textbf{Sample Mean} & \textbf{39.1\%} & \textbf{23.5\%} & \textbf{31.0\%} & \textbf{43.3\%} \\
\bottomrule
\end{tabular}}
\footnotesize{\textit{Notes: All variables from ACS 5-year estimates, 2019. Metropolitan areas sorted by mean log(1 + CLC) ascending. HO Rate = homeownership rate. White NH = White non-Hispanic. Values are tract-level means averaged within each metro.}}
\end{table}

\begin{figure} [H]
    \centering 
    \includegraphics[width=13cm]{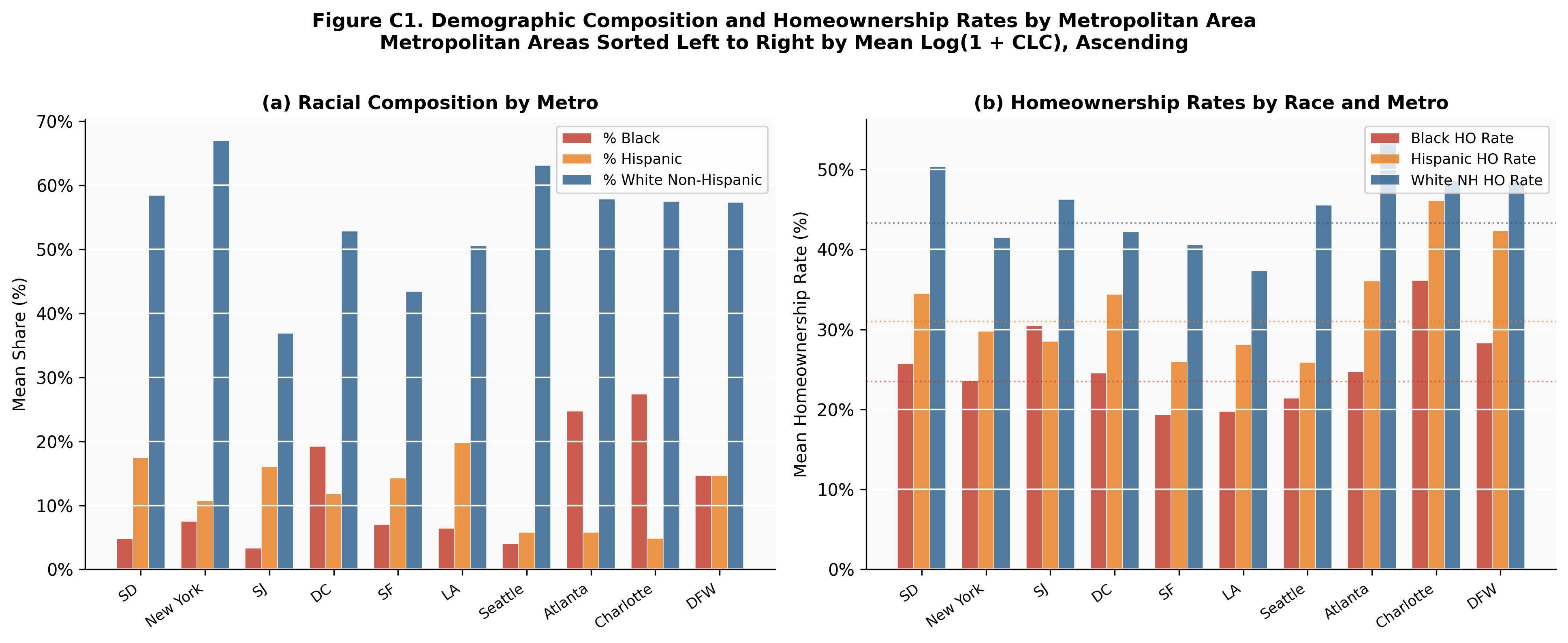}
    \caption{Panel (a): mean racial composition (\% Black, \% Hispanic, \% White non-Hispanic) by metropolitan area. Panel (b): mean homeownership rates by race and metropolitan area. Dashed horizontal lines in panel (b) mark cross-sample means for each racial group. Metropolitan areas sorted left to right by mean log(1 + CLC) ascending. Source: ACS 5-year estimates, 2019.}
    \label{fig:my_centered_image812}
\end{figure}

\end{document}